\documentstyle[prb,aps,epsf]{revtex}
\draft

\begin{document}

\twocolumn[\hsize\textwidth\columnwidth\hsize
     \csname @twocolumnfalse\endcsname

\vspace{-1.7cm}
\vspace{+0.7cm}

\title{Electrical transport properties of bulk Ni$_{c}$Fe$_{1-c}$ alloys
       and related spin-valve systems}
\author{C. Blaas}
\address{Center for Computational Materials Science,
         Technische Universit\"{a}t Wien,
         Getreidemarkt 9/158, 1060 Vienna, Austria}
\author{L. Szunyogh}
\address{Center for Computational Materials Science,
         Technische Universit\"{a}t Wien,
         Getreidemarkt 9/158, 1060 Vienna, Austria \\
         and Department of Theoretical Physics,
         Budapest University of Technology and Economics, \\
         Budafoki \'{u}t. 8, 1521 Budapest, Hungary}
\author{P. Weinberger}
\address{Center for Computational Materials Science,
         Technische Universit\"{a}t Wien,
         Getreidemarkt 9/158, 1060 Vienna, Austria}
\author{C. Sommers}
\address{Laboratoire de Physique des Solides,
         Universit\'{e} de Paris-Sud,
         91405 Orsay Cedex, France}
\author{P. M. Levy}
\address{Department of Physics,
         New York University,
         4 Washington Place, New York 10003, USA}

\date{submitted to Phys. Rev. B, July 31, 2000}
\maketitle

\begin{abstract}
Within the Kubo-Greenwood formalism we use the fully relativistic,
spin-polarized, screened Korringa-Kohn-Rostoker method together with the
coherent-potential approximation for layered systems  to calculate the
resistivity for the permalloy series Ni$_{c}$Fe$_{1-c}$. We are able to
reproduce the variation of the resistivity across the entire series;
notably the discontinuous behavior in the vicinity of the structural phase
transition from bcc to fcc. The absolute values for the resistivity are within
a factor of two of the experimental data. Also the giant magnetoresistance of
a series of permalloy-based spin-valve structures is estimated; we are able
to reproduce the trends and values observed on prototypical spin-valve
structures.
\end{abstract}


\vskip2pc]

\section{Introduction}

Permalloy is perhaps the most commonly used magnetic material in a variety
of devices. This arises primarily from its magnetic properties which combine
a reasonably high magnetic moment with a low coercivity. In many of its
applications one passes a current through the material, so that its electrical
transport properties are also important. There have been several calculations
of the transport properties of permalloy, but none has been completely
successful. Therefore, one of the outstanding challenges has been an
{\em ab initio} calculation of the electrical resistivity and anisotropic
magnetoresistance across the entire Ni$_{c}$Fe$_{1-c}$ series.
Here we present the first part which addresses the resistivity of bulk
permalloy and also the giant magnetoresistance of spin-valve structures
containing layers of permalloy. As the lattice structure changes from bcc
to fcc at $c \simeq 35\%$ we present results for both lattice structures.
We are able to reproduce the overall variation of the resistivity across the
series; in particular we find the signature changes in resistivity about
$c \simeq 35\%$ as the lattice structure changes from bcc to fcc. Remarkably
we find the magnitude of the resistivity is within a factor of two of the
experimental values; noteworthy our calculated values are larger, so that when
one relaxes the approximations we made one reduces the resistivities and
approaches the measured ones.

Here we outline our {\em ab initio} calculation and present our results for
the resistivity across the entire Ni$_{c}$Fe$_{1-c}$ series. Then we conclude
with results on prototypical spin-valve structures containing permalloy layers,
e.g., Ni$_{c}$Fe$_{1-c}$(100{\AA}) / Co(6{\AA}) / Cu(9{\AA}) / Co(6{\AA}) /
Ni$_{c}$Fe$_{1-c}$(5{\AA}). Here we find current in the plane of the layers
(CIP) magnetoresistance ratios that compare favorably with those observed in
these structures.

\section{Method of calculation}

In terms of the Kubo-Greenwood approach for disordered layered systems with
growth direction along the surface normal ($z$-axis) the in-plane conductivity
is given by \cite{jphys} 
\begin{equation}
\begin{array}{l}
\sigma_{\mu\mu}(n;{\bf c};{\cal C}) =
\displaystyle\sum_{i,j=1}^{n}\sigma_{\mu\mu}^{ij}({\bf c};{\cal C}) ,
\ \hfill \mu\in\{x,y\} , \\ \\
\sigma_{\mu\mu}^{ij} = \frac{\hbar}{\pi N_{0}\Omega_{{\rm at}}}
{\rm Tr} \left\langle J_{\mu}^{i} {\rm Im}G^{+}(\epsilon_{F})
                      J_{\mu}^{j} {\rm Im}G^{+}(\epsilon_{F}) \right\rangle ,
\end{array}
\label{e-1}
\end{equation}
where $n$ denotes the number of layers considered,
${\bf c}=\{c_{1},c_{2},\ldots ,c_{n}\}$ is a set containing the layer-wise
compositions, e.g., the layer-wise concentrations in an inhomogeneously
disordered binary alloy system A$_{c_{i}}$B$_{1-c_{i}}$ (in a homogeneously
disordered alloy $c_{i}=c, \forall i$; whereas in a spin-valve system
A$_{m}$ / B$_{n}$ / A$_{m}$ the $c_{i}=1, \forall i \in {\rm A}_{m}$,
and $c_{i}=0, \forall i \in {\rm B}_{n}$).
In principle the magnetization in each layer of a multilayered structure can
be different, and ${\cal C}=\{{\bf M}_{1},{\bf M}_{2},\ldots ,{\bf M}_{n}\}$
denotes a particular magnetic configuration, where the ${\bf M}_{i}$ refer to
the orientations of the magnetizations in the individual layers.
$\sigma_{\mu \mu }^{ij}$ is the conductivity that describes the current in
layer $i$ caused by an electric field in layer $j$, $N_{0}$ the number of
atoms per plane of atoms, $\Omega_{{\rm at}}$ is the atomic volume,
$\left\langle\cdots\right\rangle$ denotes an average over configurations,
$J_{\mu }^{i}$ is the $\mu$-th component of the current operator referenced
to the $i$-th plane, and $G^{+}$ is the electron propagator (one-particle
Green's function) from plane $i$ to $j$ at the Fermi energy $\epsilon_{F}$.
The resistivity corresponding to the conductivity given by Eq.~(\ref{e-1})
is then defined by 
\begin{equation}
\rho_{\mu\mu}(n;{\bf c};{\cal C}) = 1 / \sigma_{\mu\mu}(n;{\bf c};{\cal C}) ,
\ \quad \mu\in\{x,y\} .
\label{e-2}
\end{equation}
By assuming two different magnetic configurations, ${\cal C}$ and
${\cal C}^{\prime}$, the relative change in the resistivities between them
is given by the ratio 
\begin{equation}
R_{\mu\mu}(n;{\bf c};{\cal C};{\cal C}^{\prime}) =
\frac{\rho_{\mu\mu}(n;{\bf c};{\cal C}^{\prime})
     -\rho_{\mu\mu}(n;{\bf c};{\cal C})}
     {\rho_{\mu\mu}(n;{\bf c};{\cal C}^{\prime})}.
\label{e-2b}
\end{equation}
For example in spin-valve structures ${\cal C}^{\prime}$ and ${\cal C}$
refer to ``antiparallel'' and ``parallel'' alignments of the magnetizations
of the magnetic slabs, and this ratio is referred to as the
giant magnetoresistance (GMR) ratio.
The ratio in Eq.~(\ref{e-2b}) also serves as a definition for the
anisotropic magnetoresistance of homogeneously disordered bulk alloys,
$c_{i}=c, \forall i$, with uniform magnetization,
${\bf M}_{i}={\bf M}, \forall i$, with ${\cal C}^{\prime}$ and ${\cal C}$
referring to configurations with the magnetization pointing uniformally
parallel and perpendicular to the current axis (however, in this case
the ratio is usually taken with respect to an averaged resistivity).

In the present paper all calculations are based on self-consistent effective
potentials and effective exchange fields as obtained previously \cite{nife}
by using the fully relativistic, spin-polarized, screened Korringa-Kohn-Rostoker
method together with the coherent-potential approximation for layered systems.
For permalloy a fully relativistic spin-polarized calculation of the
resistivity is essential in order to properly account for the spin-orbit
coupling of the electron states as well as for the scattering by the
disordered constituents of these binary alloys. \cite{banhart-1,banhart-2}
The surface-Brillouin-zone integrals needed in the evaluation of the electrical
conductivity within the Kubo-Greenwood approach \cite{jphys} were obtained by
considering 1830~${\bf k}_{\parallel}$-points in the irreducible wedge
of the surface Brillouin zone. All scattering channels up to and including
$\ell_{\max}=2$ were taken into account.
Here we have chosen the magnetization to be uniformly perpendicular to the
planes of atoms, i.e., the magnetization points along the $z$-axis.
With this choice the $xx$- and $yy$-components in the above equations
are identical, hence we have taken the direction of the current
(electric field) to be along the $x$-axis so as to simulate the CIP transport
geometry (with the magnetization pointing uniformally perpendicular to the
current axis).

Consider the following systems:
\begin{equation}
\begin{tabular}{llrcl}
(a) && $\text{Ni}_{c}\text{Fe}_{1-c}(100)$ /
     & $\left( \text{Ni}_{c}\text{Fe}_{1-c}\right) _{n}$ /
     & $\text{Ni}_{c}\text{Fe}_{1-c}(100)$ \\ \\
(b) && $\text{Ni}_{c}\text{Fe}_{1-c}(100)$ /
     & $\left( \text{Ni}_{c}\text{Fe}_{1-c}\right) _{n}$ /
     & $\text{Vacuum}$ \\ \\
(c) && $\text{Vacuum}$ /
     & $\left( \text{Ni}_{c}\text{Fe}_{1-c}\right) _{n}$ /
     & $\text{Vacuum}$ .
\end{tabular}
\label{add}
\end{equation}
All three cases refer to the in-plane conductivity for a film with $n$
monolayers of permalloy; what distinguishes them are the boundary conditions.
\cite{camblong} In case (a) we have outgoing boundary conditions on both sides
as there is nothing that separates the film from the substrate it is deposited
on nor from the overlayer that caps it, i.e., some electrons will leak out into
the semi-infinite leads and their momentum information will be lost hence
producing an additional contribution to the resistivity. In (b) we have a
free surface on one side which makes it reflecting (the workfunction is finite
so that the potential barrier is not infinitely high and there will be a slight
leaking out of electrons into the vacuum that has to be taken into account).
In the third case (c) we have a perfectly flat free standing film with
reflecting boundary conditions on both sides. Only in (c) the calculated
in-plane conductivities/resistivities for a perfectly flat film are principally
independent of the film thickness $n$, at least for $n > 12$; roughness would
act like an outgoing boundary condition. However, for all three cases one
obtains the so-called bulk residual resistivity $\rho^{0}({\bf c};{\cal C})$ of
substitutionally disordered binary alloys with respect to a particular magnetic
configuration by taking the infinite-thickness limit \cite{prb-blaas}
\begin{equation}
\rho^{0}({\bf c};{\cal C}) =
\lim_{n\rightarrow\infty} \rho_{\mu\mu}(n;{\bf c};{\cal C}).
\label{e-3}
\end{equation}
This infinite-thickness limit ($n \rightarrow \infty$) can be obtained easily
since $n \rho_{\mu\mu}(n;{\bf c};{\cal C})$ becomes linear in $n$, if $n$ is
large enough; its slope simply refers to the extrapolated resistivity (see also
the discussion of Fig.~2 in the next Section).

In addition, for computational purposes a finite imaginary part $\delta$ to the
Fermi energy has to be used in the calculation of the conductivity
$\sigma_{\mu\mu}(n;{\bf c};{\cal C};\epsilon_{F}+i\delta )$. \cite{prb-blaas}
To designate resistivities calculated with finite $\delta$ we adopt the
notation 
\begin{equation}
\rho_{\mu\mu}(n;{\bf c};{\cal C};\delta ) =
1 / \sigma_{\mu\mu}(n;{\bf c};{\cal C};\epsilon_{F}+i\delta ) ,
\ \quad \mu\in\{x,y\} .
\label{e-5a}
\end{equation}
The actual resisitivity is arrived at by numerically taking the limit as the
imaginary part goes to zero 
\begin{equation}
\rho_{\mu\mu}(n;{\bf c};{\cal C}) = \lim_{\delta \rightarrow 0}
\rho_{\mu\mu}(n;{\bf c};{\cal C};\delta ) .
\label{e-5}
\end{equation}
If we are interested in the bulk residual resistivity the
($\delta \rightarrow 0$) extrapolation is simplified by the fact that the slope
of $n \rho_{\mu\mu}(n;{\bf c};{\cal C};\delta )$ (determined at large enough
$n$) is itself linear in $\delta$.

As a check on the above extrapolation procedure we used to arrive at bulk
values of the resistivity from those on films of finite thickness
(Eq.~(\ref{e-3})) we have calculated in-plane resistivities (with the
magnetization pointing uniformly along the $z$-axis) for 30 monolayers of
permalloy with a homogeneous Ni concentration of $c=0.85$ for a fixed
$\delta = 2$~mRyd with different boundary conditions:
For the reflecting boundary conditions (case (c)) we find a resistivity of
$\rho_{xx}^{\rm ref}(30;0.85;{\hat{\bf z}};2) = 20.6$~$\mu\Omega$cm.
A similar calculation with outgoing boundary conditions (case (a)) yields
$\rho_{xx}^{\rm out}(30;0.85;{\hat{\bf z}};2) = 31.5$~$\mu\Omega$cm
which by taking $n \rightarrow \infty$ yields
$\rho_{xx}^{\rm out}(\infty ;0.85;{\hat{\bf z}};2) = 20.6$~$\mu\Omega$cm;
precisely the same value as found for the reflecting boundary conditions
(case (c)). This unequivocally demonstrates that the procedure outlined above
is able to remove the resistivity that arises from the boundary conditions on
finite structures. \cite{camblong}
In Fig.~1 we show as an illustration the variation of the layer-wise
conductivities
\begin{equation}
\sigma_{xx}^{i}(n;{\bf c};{\cal C};\epsilon_{F}+i\delta ) =
\sum_{j=1}^{n}\sigma_{xx}^{ij}(n;{\bf c};{\cal C};\epsilon_{F}+i\delta ) ,
\label{e-6}
\end{equation}
across the film with reflecting and outgoing boundary conditions.
For the outgoing boundary conditions where some electrons leak into the leads
all layer-wise conductivity contributions are substantially smaller than
for the reflecting boundary conditions, the difference becoming progressively
larger when we get closer to the outer edges of the film. When we apply
reflecting boundary conditions the layer-wise conductivities are practically
constant with small oscillations from the boundary towards the center of the
film resembling the well-known Friedel oscillations.
When we take the result for $\delta = 2$~mRyd corresponding to outgoing
boundary conditions to $n \rightarrow \infty$ (Eq.~(\ref{e-3})), and then
numerically take the limit as $\delta \rightarrow 0$ (Eq.~(\ref{e-5})), we find
$\rho^{0}(0.85;{\hat{\bf z}}) =
\rho_{xx}^{\rm out}(\infty ;0.85;{\hat{\bf z}};0) = 7.1$~$\mu\Omega$cm
which is the calculated resistivity for Ni$_{0.85}$Fe$_{0.15}$ to be compared
with experiment.
It should be noted, however, that the extrapolation procedures give an upper
limit for the resistivity since due to computational constraints our
calculations can be done only up to $n \simeq 45$ with $\delta \ge 2$~mRyd
(i.e., the slopes of $n \rho_{\mu\mu}(n;{\bf c};{\cal C};\delta )$ might be
slightly too high, and close to $\delta = 0$ there might be small deviations
from linearity).
Parenthetically, a similar extrapolation procedure on a film with no disorder,
i.e., $c=1.0$, yields finite $\rho_{xx}(n;1.0;{\cal C};\delta )$ in both
cases (a) and (c) for finite $\delta$; however, the ($n \rightarrow \infty$,
$\delta \rightarrow 0$) extrapolated value
$\rho_{xx}^{\rm out}(\infty ;1.0;{\cal C};0)$ for outgoing boundary conditions,
as well as the ($\delta \rightarrow 0$) limit
$\rho_{xx}^{\rm ref}(n;1.0;{\cal C};0)$ for reflecting boundary conditions, are
both zero.

\section{Results}

\subsection{Bulk alloys}

For case (b) of Eq.~(\ref{add}), Eq.~(\ref{e-3}) is precisely the procedure
pursued in experimental studies; see Fig.~6 of McGuire and Potter \cite{ex-1}
where the thickness dependence of the resistivity of Ni$_{0.82}$Fe$_{0.18}$
films is shown. It is well known that the film resistivity increases as films
become thinner; the resistivity noticeable increases when the film thickness
becomes much smaller than the mean free path of the conduction electrons.
It is quite interesting to realize that the measured resistivity reaches its
asymptotic bulk value only for film thicknesses well above 500{\AA}.
In Fig.~2 we display the thickness dependence of our calculated
$\rho_{xx}^{\rm out}(n;c;{\hat{\bf z}};0)$ for case (a) of Eq.~(\ref{add})
as a function of the number of layers $n$ considered in the summation in
Eq.~(\ref{e-1}). As can be seen, for small values of $n$ the corresponding
changes in $\rho_{xx}^{\rm out}(n;c;{\hat{\bf z}};0)$ are large, whereas
---just as for the experimental data--- for large enough $n$
$\rho_{xx}^{\rm out}(n;c;{\hat{\bf z}};0)$ approaches towards its asymptotic
value, i.e., the functional behavior of
$n \rho_{xx}^{\rm out}(n;c;{\hat{\bf z}};0)$ becomes linear in $n$.
Inspecting our ($n \rightarrow \infty$) extrapolation we find (again in
accordance with experiment) that above 585{\AA} ($\hat{=}$ 330 monolayers)
the deviations from the bulk residual resistivity are smaller than
1~$\mu\Omega$cm, i.e., in order to talk about ``bulk'' in terms of
resistivities one has to consider a tremendously thick film of several
thousands~{\AA}.
It should be noted that in order to arrive at bulk resistivities the same kind
of linear extrapolation procedure to very large $n$ (Eq.~(\ref{e-3})) is done
both with experimental data and with our calculated resistivities
(Eqs.~(\ref{e-1}) and (\ref{e-2})).

In Fig.~3 we show our theoretical asymptotic bulk resistivities
$\rho^{0}(c;{\hat{\bf z}}) = \rho_{xx}^{\rm out}(\infty ;c;{\hat{\bf z}};0)$
obtained by performing both limits ($n \rightarrow \infty$ and
$\delta \rightarrow 0$) for the entire concentration range together with
low temperature (4.2~K) experimental data \cite{ex-1,ex-2,ex-3} for
Ni concentrations above 70\%.
The overall trends of measured data can be obtained from an inspection of
Fig.~1 of the section on Ni$_{c}$Fe$_{1-c}$ in Ref.~\onlinecite{ex-4} which
shows the temperature dependence of the resistivity for the whole range of
concentrations. From this figure one can see that by lowering the temperature
the electrical resistivity becomes increasingly sensitive to the structural
phase transition from bcc to fcc; at the temperature of liquid nitrogen
($-$195~C, the lowest isothermal curve displayed) in the Fe-rich bcc
$\alpha$-phase the resistivity reaches a maximum at about 15\% of Ni and then
slowly decreases up to 30\% of Ni, whereas in the Ni-rich fcc $\gamma$-phase
the resistivity starts to grow below 50\% of Ni and seems to diverge near the
critical concentration for the structural phase transition. In the $-$195~C
curve we also discern a weak shoulder at about 85\% Ni; this feature is well
resolved in the low temperature (4.2~K) experimental data. \cite{ex-1,ex-2,ex-3}
As can be seen from Fig.~3 our theoretical values reproduce very well the
vicinity of the structural phase transition from bcc to fcc at about 35\% of
Ni. There is indeed a maximum in the bcc phase at about 15\% of Ni; the onset
of the phase transition is clearly visible. Also ---just as the experimental
data--- in the fcc phase the theoretical values show a kind of shoulder at
about 85\% of Ni.

For this concentration ($c=0.85$) we performed an additional calculation
using a maximum angular momentum quantum number $\ell_{\max}=3$ (see also
Ref.~\onlinecite{banhart-3}). As can be seen from Fig.~3, the difference
between the $\ell_{\max}=3$ and the $\ell_{\max}=2$ (that is used in all
other calculations) results is rather small; however the trend for the
inclusion of higher scattering channels is to lower (slightly) the resistivity.

Although the theoretical resistivities agree rather well with the experimental
ones, the difference at a particular concentration being of the order of
3~$\mu\Omega$cm, it seems appropriate to comment on the putative sources of
the difference.
First of all, in the present use of the Kubo-Greenwood equation for layered
systems no vertex corrections arising from the configurational average of the
product of two Green's functions are included (see also the discussion in
Ref.~\onlinecite{jphys}). In another calculation \cite{banhart-1,banhart-2}
of the resistivity of bulk Ni$_{c}$Fe$_{1-c}$ the vertex corrections were
calculated and found to be quite small over the entire range of concentrations,
\cite{vernes} i.e., no larger than the differences between the $\ell_{\max}=2$
and $\ell_{\max}=3$ results at $c=0.85$ in Fig.~3.
Second, the single site approximation to the coherent-potential approximation
is used to describe the electronic structure of substitutional alloys. This in
turn implies that short range order or concentration fluctuations are excluded.
Both the inclusion of vertex corrections and the inclusion of short range order
will reduce the present theoretical resistivity values. Furthermore, in
magnetic alloys magnetic ordering or clustering can occur; as it implies some
ordering, which we have not taken into account, it will again (slightly) lower
the resistivity.

In two previous theoretical papers \cite{banhart-1,banhart-2} the validity
of the so-called two-current model for transport in ferromagnetic systems
with strong spin-dependent disorder was tested; the resistivity and the
anisotropic magnetoresistance for Ni$_{c}$Fe$_{1-c}$ bulk alloys was calculated
by solving the Kubo-Greenwood equation using a spin-polarized, relativistic
version of the Korringa-Kohn-Rostoker method together with the
coherent-potential approximation. As these calculations were done on a fcc
lattice for the whole concentration range their results are not applicable to
the Fe-rich bcc alloys. Nonetheless a very weak shoulder at about 80\% Ni can
be discerned from their Fig.~4 \cite{banhart-1} ---in agreement with the
experimental data and our present calculation, see Fig.~3. However, the
resistivities they find for 80\% Ni, $\rho \simeq 1.8~\mu\Omega$cm, as well as
the maximum at about 40\% Ni, $\rho \simeq 3.2~\mu\Omega$cm, are too low by
about a factor of two as compared to available experimental data.

\subsection{Spin-valve structures}

The main purpose of our study has been to show how well resistivities for
disordered magnetic systems with only two-dimensional translational symmetry
(films) can be described, and to confirm that by taking the appropriate
limit we derive the bulk resistivities from those calculated for films.
Since a large number of spin-valve structures that are used in applications,
e.g., the read heads of hard disk drives, contain permalloy layers, we
conclude with some results on their transport properties. Including permalloy
layers into a multilayer system the electronic structure and the magnetic
properties are considerably more complicated than for bulk permalloy;
specifically they are sensitive to the thickness and Ni concentration $c$ of
the permalloy layers, and to the constituency of the neighboring layers.
Therefore for concreteness we focus on a permalloy-based system that has been
studied, e.g., Si-substrate / Cu-contacts(1500{\AA}) /
Ni$_{c}$Fe$_{1-c}$(100{\AA}) / Co(6{\AA}) / Cu(9{\AA}) / Co(6{\AA}) /
Ni$_{c}$Fe$_{1-c}$(5{\AA}) / NiO(250{\AA}), \cite{NRL} and we replace it with
the reference system $\text{Ni}_{c}\text{Fe}_{1-c}(100)$ /
$\left( \text{Ni}_{c}\text{Fe}_{1-c}\right) _{12}$
Co$_{4}$ Cu$_{n}$ Co$_{4}$
$\left( \text{Ni}_{c}\text{Fe}_{1-c}\right) _{3}$ / Vacuum.
In all cases shown a fcc(100) stacking was chosen, the lattice spacing refers
to that of the corresponding fcc bulk Ni$_{c}$Fe$_{1-c}$ system. In this
transcription the rather thick permalloy layer of about 100{\AA}
($\hat{=}$ 56 monolayers) is considered as a substrate, the insulating NiO
layer is replaced by a free surface (vacuum), and all other thicknesses are
expressed (in monolayers) as close as possible to the measured structure. The
12 layers of Ni$_{c}$Fe$_{1-c}$ are deliberately chosen so as to guarantee a
smooth matching to the substrate via a selfconsistent calculation; this in
turn would also allow us to change the left boundary condition such that in
principle a system of the type
Vacuum / $\left( \text{Ni}_{c}\text{Fe}_{1-c}\right) _{56}$
Co$_{4}$ Cu$_{n}$ Co$_{4}$
$\left( \text{Ni}_{c}\text{Fe}_{1-c}\right) _{3}$ / Vacuum
could be described by assuming that the layers in the very interior of the
thick Ni$_{c}$Fe$_{1-c}$ layer exhibit bulk-like properties.

We have considered two series: in the first the Cu-spacer thickness is
confined to $n=5$ and we vary the Ni concentration $c$ while maintaining an
fcc lattice throughout; in the second we set $c=0.85$ and vary the number of
Cu-spacer layers $n$. While the structure of Ni$_{c}$Fe$_{1-c}$ bulk alloys is
bcc for $c<35\%$ when there are relatively thin layers in a fcc multilayered
structure it is appropriate to maintain the fcc lattice structure even for
the Fe-rich alloys. In the parallel configuration of the spin valve the
magnetizations of all permalloy and Co layers are aligned uniformly
perpendicular to the plane of the layers; in the antiparallel configuration
they are still normal to the layers but the magnetizations of the two magnetic
slabs (separated by the Cu-spacer) point in opposite directions to one another.
The conductivity is evaluated at a finite imaginary part to the Fermi energy,
$\delta = 2$~mRyd; this is necessary for otherwise the calculations would
take an inordinate time. In principle, in order to obtain the resistivity we
have to take the limit as $\delta \rightarrow 0$ (see Eq.~(\ref{e-5})).
However, it is virtually impossible for technical reasons to do this up till
now for a whole series of spin-valve structures; therefore we do not show the
resistivities we calculated for finite $\delta$. But with the assumption that
both resistivities for the parallel and the antiparallel configuration of the
spin valve scale uniformly with $\delta$, and therefore cancel out in the
ratio $R$ (Eq.~(\ref{e-2b})), we have used them to estimate the CIP-MR ratios
that one can expect from these spin-valve structures.

In Fig.~4a we show the variation of the calculated CIP-MR ratio of the reference
system with the Ni concentration $c$ for the fixed Cu-spacer thickness $n=5$.
As can be seen the CIP-MR varies only very little (between about 17.5--21.5\%)
over the whole range of concentrations considered. In Fig.~4b we show the
variation of the CIP-MR ratio for the reference system corresponding to $c=0.85$
with respect to the number of Cu-spacer layers $n$. By doubling the number of
Cu-spacer layers, i.e., by going from $n=4$ to $n=8$, the CIP-MR is reduced by
almost a factor of two.
This dramatic decrease can be understood in terms of a simple model:
since on increasing the thickness of the nonmagnetic spacer layer $t_{\rm nm}$
the CIP-MR decreases as $ (1/t_{\rm nm}) \exp(-t_{\rm nm}/\lambda_{\rm nm})$,
\cite{levy} where $\lambda_{\rm nm}$ is the mean free path of the conduction
electrons in the nonmagnetic spacer layer, the ratio
$R_{xx}(t_{\rm nm})/R_{xx}(2t_{\rm nm}) \approx 2$ for
$t_{\rm nm} \ll \lambda_{\rm nm}$ (which is a necessary condition to find
at all a CIP-MR and certainly valid for our $t_{\rm nm}$ varying between
7.1~{\AA} and 14.2~{\AA}).
Our findings seem to be confirmed by experiments made on similar systems
\cite{MOTOROLA} albeit at room temperature, while our calculated values are for
$T=0$~K. Depending on the actual thicknesses of: the bottom and top
Ni$_{c}$Fe$_{1-c}$ layers, of the Co slabs and of the Cu spacer, and the
pinning and capping layers used, the experimental CIP-MR ratios in
permalloy-related spin-valve systems have been found to be in the range between
10 and 20\%. Fig.~4 nicely shows that this is about the range that our
theoretical calculations predict when we use the resistivities calculated with
finite $\delta$, and when we make the assumption that for both the parallel and
the antiparallel configurations of the spin valve the resistivity scales
uniformly with $\delta$ and therefore cancels out in the CIP-MR ratio $R$.

\section{Summary}

By using our fully relativistic, spin-polarized, screened Korringa-Kohn-Rostoker
method together with the coherent-potential approximation for layered systems
we have calculated the resistivity across the entire Ni$_{c}$Fe$_{1-c}$
(permalloy) series of bulk alloys within the Kubo-Greenwood approach.
We are able to reproduce the overall variation of the resistivity as a function
of the Ni concentration; in particular we reproduce the discontinuity in the
vicinity of the structural change from bcc to fcc. Also we obtain very
reasonable absolute values for the resistivities. We have also estimated the
CIP-MR ratios for permalloy-based spin-valve structures and within the limits
of our current calculation find ratios in reasonably good agreement with those
measured on prototypical spin valves; there is of course the caveat that most
of the data on spin valves is quoted at room temperature while our results are
for $T=0$~K. It should be stressed that our results are based on
{\em ab initio} calculations with no adjustable parameters. Viewed in this
light the calculated resistivities and GMR ratios are noteworthy.

\acknowledgements

This work was supported by the Austrian Science Foundation (T27-TPH, P12146),
the Hungarian National Science Foundation (OTKA T030240, T029813),
the Defense Advanced Research Projects Agency and the Office of Naval
Research (N00014-96-1-1207, MDA 972-99-C-009), NATO (CRG 960340),
and the TMR network (EMRX-CT96-0089). We also wish to thank the CNRS IDRIS
Computing Center at Orsay for calculations done on their T3E Cray.

\pagebreak
\mbox{} \vspace{-2cm}

\begin{figure}
\epsfxsize=7cm \centerline{\epsffile{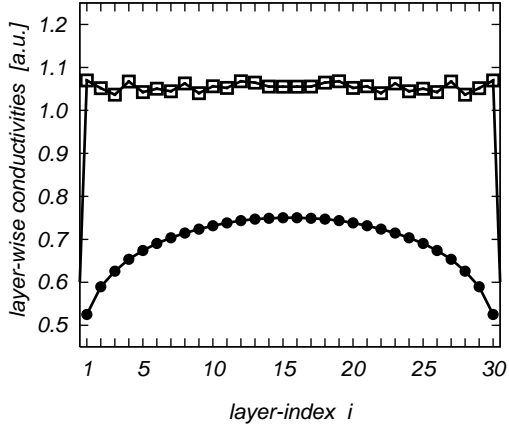}} \vspace{0.3cm}
\caption{
Variation of the layer-wise conductivities
$\sigma_{xx}^{i}(30;0.85;{\hat{\bf z}};\epsilon_{F}+i2)$
(Eq.~(\protect\ref{e-6})) across the film for 30 monolayers of permalloy with a
homogeneous Ni concentration of $c=0.85$ for a fixed $\delta = 2$~mRyd with
different boundary conditions: The values for the reflecting boundary
conditions are show as empty squares, those for the outgoing boundary
conditions as full circles.}
\end{figure}

\begin{figure}
\epsfxsize=7cm \centerline{\epsffile{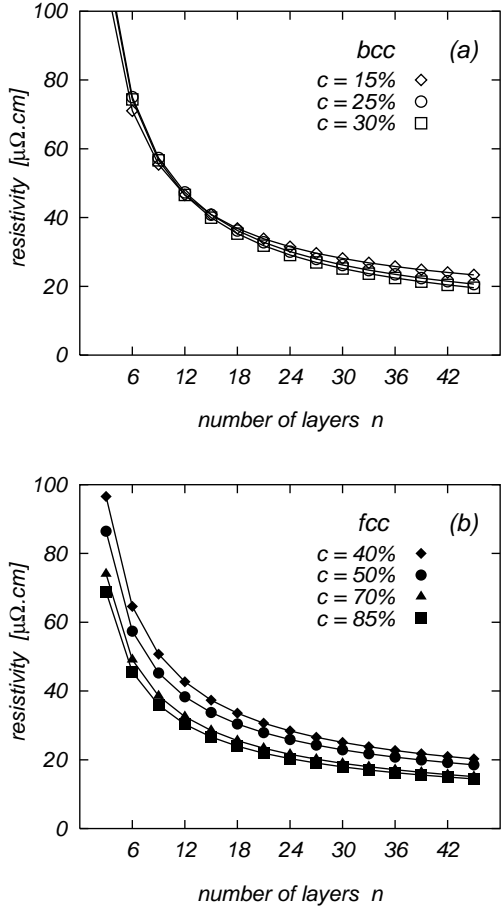}} \vspace{0.3cm}
\caption{Thickness dependence of the current-in-plane resistivity in bcc (a)
and fcc (b) $\text{Ni}_{c}\text{Fe}_{1-c}(100)$ / 
$\left( \text{Ni}_{c}\text{Fe}_{1-c}\right) _{n}$
/ $\text{Ni}_{c}\text{Fe}_{1-c}(100)$ alloys
($\rho_{xx}^{\rm out}(n;c;{\hat{\bf z}};0)$).
The Ni concentration in \% is indicated explicitly, $n$ denotes the number of
layers considered in the summation in Eq.~(\protect\ref{e-1}).}
\end{figure}

\pagebreak
\mbox{} \vspace{-2cm}

\begin{figure}
\epsfxsize=7cm \centerline{\epsffile{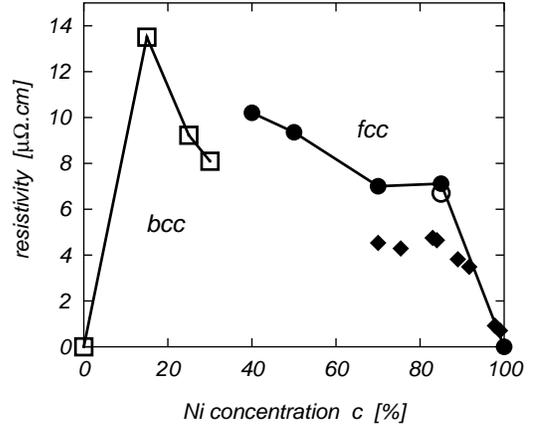}} \vspace{0.3cm}
\caption{Concentration dependence of the current-in-plane resistivity of
``bulk'' Ni$_{c}$Fe$_{1-c}$ alloys calculated as the infinite-thickness limit
of $\text{Ni}_{c}\text{Fe}_{1-c}(100)$ / 
$\left( \text{Ni}_{c}\text{Fe}_{1-c}\right) _{n\rightarrow\infty}$
/ $\text{Ni}_{c}\text{Fe}_{1-c}(100)$
($\rho^{0}(c;{\hat{\bf z}}) = \rho_{xx}^{\rm out}(\infty ;c;{\hat{\bf z}};0)$).
The results for the bcc $\alpha$-phase are shown as empty squares,
for the fcc $\gamma$-phase as full circles, the open circle at $c=85\%$ refers
to a calculation with $\ell_{\max}=3$. Low temperature (4.2~K) experimental
values displayed by full diamonds are taken from
Refs.~\protect\onlinecite{ex-1,ex-2,ex-3}.}
\end{figure}

\begin{figure}
\epsfxsize=7cm \centerline{\epsffile{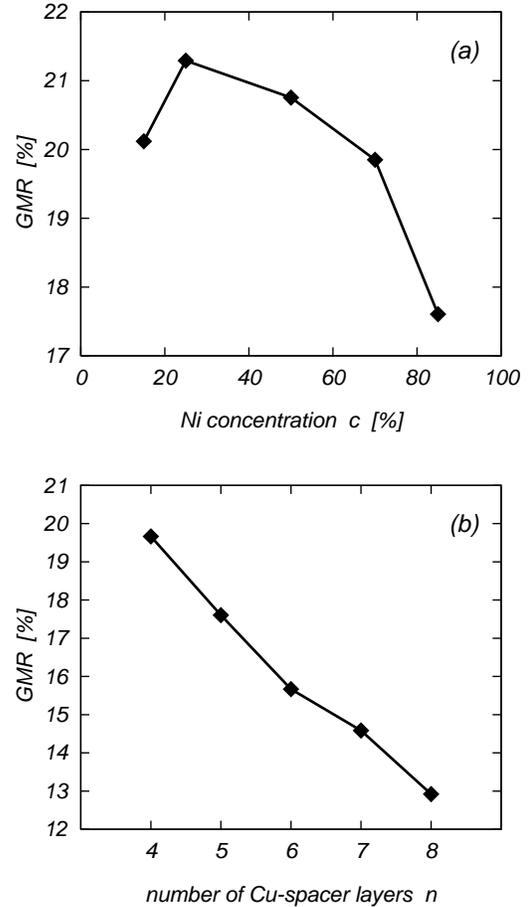}} \vspace{0.3cm}
\caption{Current-in-plane magnetoresistance for permalloy-based spin valves:
(a) Concentration dependence for fcc-based
$\text{Ni}_{c}\text{Fe}_{1-c}(100)$ / 
$\left( \text{Ni}_{c}\text{Fe}_{1-c}\right) _{12}$ 
Co$_{4}$ Cu$_{5}$ Co$_{4}$
$\left( \text{Ni}_{c}\text{Fe}_{1-c}\right) _{3}$ / Vacuum systems.
(b) Dependence on the number of Cu-spacer layers $n$ in fcc-based
$\text{Ni}_{0.85}\text{Fe}_{0.15}(100)$ / 
$\left( \text{Ni}_{0.85}\text{Fe}_{0.15}\right) _{12}$ 
Co$_{4}$ Cu$_{n}$ Co$_{4}$
$\left( \text{Ni}_{0.85}\text{Fe}_{0.15}\right) _{3}$ / Vacuum.
Experimental values are found to be in the range
10--20\%.\protect\cite{NRL,MOTOROLA}}
\end{figure}

\end{document}